\documentclass[aps,twocolumn,preprintnumbers,amsmath,amssymb,nofootinbib,superscriptaddress,notitlepage]{revtex4}

\usepackage{epsfig}
\usepackage{slashed}

\usepackage{multirow}

\begin{document}

\title{The $\bar{D}^* \bar{D}^* \Sigma_c$ Three-Body System}

	\author{Manuel Pavon Valderrama}\email{mpavon@buaa.edu.cn}
\affiliation{School of Physics and Nuclear Energy Engineering, \\
International Research Center for Nuclei and Particles in the Cosmos and \\
Beijing Key Laboratory of Advanced Nuclear Materials and Physics, \\
Beihang University, Beijing 100191, China} 

\date{\today}

%\date{\today}

\begin{abstract} 
  \rule{0ex}{3ex}
  In the molecular picture the hidden-charm, pentaquark-like $P_c(4450)$
  resonance is a $\bar{D}^* \Sigma_c$ bound state with
  quantum numbers $I=\tfrac{1}{2}$ and $J^P = \tfrac{3}{2}^-$.
  If this happens to be the case, it will be natural to expect
  the existence of $\bar{D}^* \bar{D}^* \Sigma_c$ three-body bound states.
  The most probable quantum numbers for a bound $\bar{D}^* \bar{D}^* \Sigma_c$
  trimer are the isoscalar $J^P = \tfrac{1}{2}^+$, $\tfrac{5}{2}^+$
  and the isovector $J^P = \tfrac{3}{2}^+$, $\tfrac{5}{2}^+$ configurations.
  Calculations within a contact-range theory indicate a trimer
  binding energy $B_3 \sim 3-5\,{\rm MeV}$ and $14-16\,{\rm MeV}$
  for the isoscalar $\tfrac{1}{2}^+$ and $\tfrac{5}{2}^+$ states
  and $B_3 \sim 1-3\,{\rm MeV}$ and $3-5\,{\rm MeV}$
  for the isovector $\tfrac{3}{2}^+$ and $\tfrac{5}{2}^+$ states, respectively,
  with $B_3$ relative to the $\bar{D}^* P_c(4450)$ threshold.
  These predictions are affected by a series of error sources
  that we discuss in detail.
\end{abstract}

\maketitle

\section{Introduction}

The discovery of two hidden charm pentaquark-like resonances
by the LHCb~\cite{Aaij:2015tga}, the $P_c(4380)$ and $P_c(4450)$,
begs the question of what is their nature.
These resonances are indeed different:
while the $P_c(4380)$ is relatively broad, the $P_c(4450)$
is narrow and is located close to a few meson-baryon thresholds,
namely the $\bar{D}\Lambda_{c}(2595)$, $\bar{D}^*\Sigma_c$,
$\bar{D}^*\Sigma_c^*$ and $J/\psi \, p$ thresholds.
This coincidence makes the $P_c(4450)$ a good candidate for a hadronic molecule,
a type of exotic hadron theorized a few decades
ago~\cite{Voloshin:1976ap,DeRujula:1976qd}.
In particular the $P_c(4450)$ has been suggested to be a
$\bar{D}^* \Sigma_c$~\cite{Roca:2015dva,He:2015cea,Xiao:2015fia},
a $\bar{D}^* \Sigma_c^*$ molecule~\cite{Chen:2015loa,Chen:2015moa}
(in these two cases probably with a small
admixture of $\bar{D} \Lambda_{c}(2595)$~\cite{Burns:2015dwa,Geng:2017hxc}),
and a $\chi_{c1} \, p$ molecule~\cite{Meissner:2015mza}.
Of these possibilities, the most natural one probably is $\bar{D}^* \Sigma_c$.
First, the interaction between a heavy baryon and antimeson is expected
to be mediated by the exchange of light mesons, providing
a mechanism to justify the existence of attraction.
Second ,the binding energy of the $\Sigma_c \bar{D}^*$,
$B_2 = 12 \pm 3 \,{\rm MeV}$, translates into a bound state that is not
excessively compact, which is compatible with the molecular hypothesis.
The expected size is of the order of $1 / \sqrt{2\mu B_2} \sim 1.2\,{\rm fm}$,
with $\mu$ is the reduced mass of the molecule.
Besides the molecular hypothesis there are other competing explanations
for the nature of the $P_c^*$, which include a genuine pentaquark~\cite{Diakonov:1997mm,Jaffe:2003sg,Yuan:2012wz,Maiani:2015vwa,Lebed:2015tna,Li:2015gta,Wang:2015epa},
threshold effects~\cite{Guo:2015umn,Liu:2015fea} (for a detailed
discussion see Ref.~\cite{Bayar:2016ftu}),
baryocharmonia~\cite{Kubarovsky:2015aaa},
a molecule bound by {\it colour chemistry}~\cite{Mironov:2015ica}
or a soliton~\cite{Scoccola:2015nia}.
The molecular hypothesis relies on the quantum numbers of the pentaquark
to be $J^P = \tfrac{3}{2}^{-}$ or alternatively $\tfrac{5}{2}^-$.
Experimentally $J^P$ is not uniquely determined yet, with $\tfrac{3}{2}^{+}$
and $\tfrac{5}{2}^+$ also probable~\cite{Aaij:2015tga}.
Until the quantum numbers of the $P_c^*$ are known,
the discussion about its nature will remain theoretical.

If the $P_c(4450)$ --- the $P_c^*$ from now on --- is indeed a
$J^P=\tfrac{3}{2}^-$ $\bar{D}^* \Sigma_c$ bound state, its location
will provide useful information about the dynamics of this two-body system.
This information can be used to deduce the existence of new molecules.
For instance, from heavy quark spin symmetry it is sensible to expect
the existence of a $J^P=\tfrac{5}{2}^-$ $\bar{D}^* \Sigma_c^*$ partner of
the $P_c^*$ with a mass of $5515\,{\rm MeV}$~\cite{Liu:2018zzu}.
In this manuscript we argue that if the $P_c^*$ turns out to be molecular
the dynamics of the $\bar{D}^* \Sigma_c$ two-body system
imply the existence of a series of
$\bar{D}^* \bar{D}^* \Sigma_c$ three-body bound states.
Concrete calculations in a contact-range theory lead to the following
predictions:
\begin{enumerate}
\item[(i)] a $J=\tfrac{1}{2}$, $I=0$ trimer with $B_3 \sim 3-5 \,{\rm MeV}$,
\item[(ii)] a $J=\tfrac{3}{2}$, $I=1$ trimer with $B_3 \sim 1-3 \,{\rm MeV}$,
\item[(iii)] a $J=\tfrac{5}{2}$, $I=0$ trimer
  with $B_3 \sim 14-16 \,{\rm MeV}$,
\item[(iv)] a $J=\tfrac{5}{2}$, $I=1$ trimer with $B_3 \sim 3-5 \,{\rm MeV}$,
\end{enumerate}
where the trimer binding energy $B_3$ is relative to the hadron-dimer
threshold, i.e. the $\bar{D}^* P_c^*$ threshold.
There are a series of uncertainties related to the previous predictions
of which the most crucial one is whether the $P_c^*$ is really
a  $\bar{D}^* \Sigma_c$ bound state.
Until the nature of the $P_c^*$ is settled, these trimers will
remain a theoretical possibility.
Conversely the experimental production of these trimers
or their detection in the lattice will strongly suggest
a molecular nature for the $P_c^*$.
It is worth noticing that analogous trimer predictions have been made
for other molecular candidates. For instance, from the hypothesis that
the $X(3873)$ is a $D^*\bar{D}$ molecule it is possible to theorize
about $D^* \bar{D} K$, $D^* \bar{D}^* K$~\cite{Ren:2018pcd},
$D^* D^* \bar{D}$ and $D^* D^* \bar{D}^*$
trimers~\cite{Valderrama:2018sap}.
Within more phenomenological approaches there has also been a growing interest
in the possibility of three-body systems in the heavy sector,
for instance $D^* \bar{D}^* \rho$~\cite{Bayar:2015oea},
$B^* \bar{B}^* \rho$~\cite{Bayar:2015zba},
$B D D$, $B D \bar{D}$, $B^* D \bar{D}$, etc.~\cite{Dias:2017miz,Dias:2018iuy},
just to mention a few recent works.

The manuscript is structured as follows:
in Sect.~\ref{sec:Faddeev} we write the Faddeev equations
for the $\bar{D}^* \bar{D}^* \Sigma_c$ system.
In Sect.~\ref{sec:Efimov} we consider the $\bar{D}^* \bar{D}^* \Sigma_c$ system
in the unitary limit, i.e. when the binding energy of the $\bar{D}^* \Sigma_c$
system approaches zero.
In Sect.~\ref{sec:PcD} we present our predictions of three body states.
Finally in Sect.~\ref{sec:Conclusions} we summarize
the results of this manuscript.

\section{Faddeev Equations for the $\Sigma_c \bar{D}^* \bar{D}^*$ System}
\label{sec:Faddeev}

In this section we write the Faddeev decomposition and equations
for the $\bar{D}^* \bar{D}^* \Sigma_c$ three body system.
We will consider the S-wave three body states only, as they have the highest
chances of binding. We will make the following assumptions:
\begin{itemize}
\item[(i)] the $\bar{D} \bar{D}$, $\bar{D} \bar{D}^*$ and $\bar{D}^* \bar{D}^*$ 
pairs do not interact,
\item[(ii)]
  the $\bar{D}^* \Sigma_c$ only interacts in the $P_c^*$ channel,
\item[(iii)] the $\bar{D}^* \Sigma_c$ interaction in the $P_c^*$ channel
can be described in terms of a contact-range interaction.
\end{itemize}
This leads to a considerable simplification of the Faddeev equations.
Equivalently we can consider the $\bar{D}^* \bar{D}^* \Sigma_c$ system
from the effective field theory point of view.
In this case we will say that the $\bar{D}^* \Sigma_c$ contact-range
interaction is leading order, while all the other interactions
are subleading and can be ignored at leading order.
We will review the previous set of assumptions at the end of the manuscript.

\subsection{The Equations for $\bar{D}^* \bar{D}^* \Sigma_c$}

We begin by writing the three body wave function in terms of Faddeev
components for the $\bar{D}^{*} \bar{D}^{*} \Sigma_c$ system
\begin{eqnarray}
  \Psi_{3B} &=& \sum_{\beta}\,
  \left[ \phi_{\beta}(\vec{k}_{23},\vec{p}_1) + \zeta_{\beta}\,
    \phi_{\beta}(\vec{k}_{31},\vec{p}_2)
    \right] \, \nonumber \\ && \quad \times
  | S_{12} \otimes \frac{1}{2}  \rangle_S\,| I_{12} \otimes 1 \rangle_I 
  \, ,
\end{eqnarray}
where we have labeled the two $D^*$ mesons as particles $1$ and $2$
and the $\Sigma_c$ as particle $3$.
The summation index $\beta = (S_{12}, I_{12})$ refers to the spin and isospin
of the $\bar{D}^{*} \bar{D}^{*}$ subsystem.
The spin and isospin piece of the wave function is indicated by
\begin{eqnarray}
  | S_{12} \otimes \frac{1}{2}  \rangle_S\,| I_{12} \otimes 1 \rangle_I \, , 
\end{eqnarray}
where $S$ and $I$ are the total spin and isospin of
the $\bar{D}^* \bar{D}^* \Sigma_c$ three body system.
The sign $\zeta_{\beta} = \pm 1$ indicates whether the spatial
part of the $\bar{D}^{*} \bar{D}^{*}$ wave function
is symmetric or antisymmetric.
We will ignore the $\zeta_{\beta} = -1$ configurations: they require
the orbital angular momentum of the $12$ subsystem to be $L_{12} \geq 1$,
which implies that they are suppressed for the positive parity states
(they depend on the P-wave $\bar{D}^* \Sigma_c$ interaction).
We define the Jacobi momenta as follows
\begin{eqnarray}
  \vec{k}_{ij} &=& \frac{m_j \vec{k}_i - m_i \vec{k}_j}{m_i + m_j} \, , \\
  \vec{p}_{k} &=& \frac{1}{M_T}\,\left[ (m_i + m_j)\,\vec{k}_k -
    m_k\,(\vec{k}_i + \vec{k}_j) \right] \, , 
\end{eqnarray}
with $m_1$, $m_2$, $m_3$ the masses of particles $1$, $2$, $3$,
$M_T = m_1 + m_2 + m_3$ the total mass and $ijk$ an even permutation of $123$.
In particular we have $m_1 = m_2 = m(\bar{D}^{*})$ and $m_3 = m(\Sigma_c)$.
At this point it will be helpful to make the following observation
about the Faddeev components: in principle there are up to three
Faddeev components for each channel $\beta$.
Of these, the first two Faddeev components are symmetric or antisymmetric
(as indicated by the sign $\zeta_{\beta}$) because particles $1$ and $2$
are identical: $\phi_{\beta}(\vec{k},\vec{p})$ denotes these components.
Finally the third Faddeev component vanishes because we have considered
that the $\bar{D}^{*} \bar{D}^{*}$ subsystem does not interact.

We assume a contact-range $\bar{D}^* \Sigma_c$ potential of the type
\begin{eqnarray}
  V_{\Sigma_c \bar{D}^*} = C_{\sigma \tau} g(k) g(k') \, ,  
\end{eqnarray}
where $C_{\sigma \tau}$ is a coupling constant and
$g(p)$ is a regulator function.
The subscripts $\sigma$ and $\tau$ indicate the spin and isospin channel
of the $\bar{D}^* \Sigma_c$ subsystem,
i.e. $\sigma = J_{23}$ and $\tau = I_{23}$.
The $\bar{D}^* \Sigma_c$ T-matrix can be written as
\begin{eqnarray}
  T_{23} = t_{\sigma \tau}(Z) g(k) g(k') \, ,
\end{eqnarray}
depending on the spin and isospin channel,
where $Z$ represents the energy of
the few-body system with respect to the few-body mass threshold.
With this two-body T-matrix, the Faddeev component $\phi_{\beta}$
admits the ansatz
\begin{eqnarray}
  \phi_{\beta}(k,p) =
  \frac{g(k)}{Z - \frac{k^2}{2 \mu_{23}} - \frac{p^2}{2 \mu_1}} a_{\beta}(p)
  \, ,
\end{eqnarray}
with the reduced masses defined as
\begin{eqnarray}
  \frac{1}{\mu_{ij}} &=& \frac{1}{m_i} + \frac{1}{m_j} \, , \label{eq:mu_ij} \\
  \frac{1}{\mu_{k}} &=& \frac{1}{m_k} + \frac{1}{m_i + m_j} \label{eq:mu_k} \, .
\end{eqnarray}
From the previous, we find that $a_{\beta}(p)$ follows the integral equation
\begin{eqnarray}
  a_{\beta}(p_1) &=&
  \left[
    \sum \lambda^{\beta \gamma}_{\sigma \tau} t_{\sigma \tau}(Z_{23})\, \right]
  \nonumber \\
  &\times&
  \int \frac{d^3 p_2}{(2\pi)^3}\,
  B^0_{12}(\vec{p}_1, \vec{p}_2)\,a_{\gamma}(p_2) \, ,
\end{eqnarray}
where $\beta$ and $\gamma$ refer to the $S_{12}$ and $I_{12}$
spin and isospin channels and
$Z_{23} = Z -\tfrac{p_1^2}{2 m_1}\,\tfrac{M_T}{m_2 + m_3}$.
The driving term $B^0_{12}$ is given by
\begin{eqnarray}
  B^0_{ij} (\vec{p}_i, \vec{p}_j) =
  \frac{g(q_i)\,g(q_j)}
       {Z - \frac{p_1^2}{2 m_1} - \frac{p_2^2}{2 m_2} - \frac{p_3^2}{2 m_3}} \, ,
\end{eqnarray}
with $\vec{p}_1 + \vec{p}_2 + \vec{p}_3 = 0$ and
\begin{eqnarray}
  \vec{q}_k = \frac{m_j \vec{p}_i - m_i \vec{p}_j}{m_j + m_i} \, .
\end{eqnarray}
The integral equation can be discretized, in which case finding
the bound state solution reduces to an eigenvalue problem.

Regarding the $\bar{D}^* \Sigma_c$ potential,
the only  $\sigma \tau$ channel for which we know
the interaction is the $P_c^*$ channel,
i.e. $\sigma = \frac{3}{2}$ and $\tau = \frac{1}{2}$.
For simplicity we will only consider the interaction in the $P_c^*$ channel
and neglect the interaction in all other channels
\begin{eqnarray}
  t_{\sigma \tau}(Z) = t_{P_c^*}(Z)\,
  \delta_{\sigma \frac{3}{2}} \delta_{\tau \frac{1}{2}} \, .
\end{eqnarray}
Equivalently, in the effective field theory language this means
that we are considering the $P_c^*$ interaction as leading order
and the interaction in all the other channels
as subleading corrections.
This point is important as it will entail a simplification
in the Faddeev equations: coupled-channel equations will
become single-channel ones.

To understand this simplification we first notice that
there are nine possible quantum numbers for the S-wave
$\bar{D}^* \bar{D}^* \Sigma_c$ system: three spin states
$J = \frac{1}{2}$, $\frac{3}{2}$, $\frac{5}{2}$
compounded by three isospin states $I=0$, $1$, $2$.
Seven of these configurations involve a single channel,
i.e. a single possible $\beta = (S_{12},I_{12})$ configuration
for the spin and isospin of the $\bar{D}^* \bar{D}^*$ subsystem.
Then there are two quantum numbers that involve coupled channels:
$J=\frac{1}{2}$, $\frac{3}{2}$ with $I=1$.
Now if we ignore all the $\bar{D}^* \Sigma_c$ interactions
with the exception of the $P_c^*$ channel,
we can simply make the substitution
\begin{equation}
  \lambda^{\beta \gamma}_{\sigma \tau} \,t_{\sigma \tau}\to
  \lambda^{\beta \gamma}_{P_c^*} \, t_{P_c^*} \, .
\end{equation}
The matrix $\lambda^{\beta \gamma}_{P_c^*}$ can be diagonalized, in which case
the coupled-channel equation reduces to a single-channel one:
\begin{eqnarray}
  a(p_1) &=&
  \lambda\, t_{P_c^*}(Z_{23})\,
%  \nonumber \\ &\times&
  \int \frac{d^3 p_2}{(2\pi)^3}\,
  B^0_{12}(\vec{p}_1, \vec{p}_2)\,a(p_2) \, , \nonumber \\
  \label{eq:PcD-trimer}
\end{eqnarray}
where $\lambda$ is one of the eigenvalues of $\lambda^{\beta \gamma}_{P_c^*}$.
In fact this is the equation that we will use in all cases to find
the binding energies of the $\bar{D}^* \bar{D}^* \Sigma_c$ trimers.
The factors $\lambda$ for the different trimer quantum numbers 
can be found in Table \ref{tab:lambda1-spin}.

\begin{table*}[!ttt]
\begin{tabular}{|cc|c|cccc|c|}
\hline\hline
$J^P$  & $I$ & $\beta = (S_{12}, I_{12})$ &
$\lambda_{\frac{1}{2},\frac{1}{2}}$ &
$\lambda_{\frac{1}{2},\frac{3}{2}}$ &
$\lambda_{\frac{3}{2},\frac{1}{2}}$ &
$\lambda_{\frac{3}{2},\frac{3}{2}}$ &
$\lambda$ \\
\hline
$\frac{1}{2}^+$ & $0$ & $(0,1)$ & $\frac{1}{3}$ & $0$ &
$\frac{2}{3}$ & $0$ & $\frac{2}{3}$ \\ 
$\frac{1}{2}^+$ & $1$ & $\{ (0,1), (1,0) \}$ &
$\left( \begin{smallmatrix} \frac{2}{9} & \frac{2}{9} \\ \frac{2}{9} & \frac{2}{9} \\ \end{smallmatrix} \right)$ &  
$\left( \begin{smallmatrix} \frac{1}{9} & -\frac{2}{9} \\ -\frac{2}{9} & \frac{1}{9} \\ \end{smallmatrix} \right)$ &
$\left( \begin{smallmatrix} \frac{4}{9} & -\frac{2}{9} \\ -\frac{2}{9} & \frac{1}{9} \\ \end{smallmatrix} \right)$ &
$\left( \begin{smallmatrix} \frac{2}{9} & \frac{2}{9} \\ \frac{2}{9} & \frac{2}{9} \\ \end{smallmatrix} \right)$ &
$\frac{5}{9}$, $0$
\\
\hline
$\frac{3}{2}^+$ & $0$ & $(2,1)$ & $\frac{5}{6}$ & $0$ &
$\frac{1}{6}$ & $0$ & $\frac{1}{6}$ \\ 
$\frac{3}{2}^+$ & $1$ &
$\{(1,0),(2,1)\}$ &
$\left( \begin{smallmatrix} \frac{1}{18} & \frac{\sqrt{10}}{18} \\ \frac{\sqrt{10}}{18} & \frac{5}{9} \\ \end{smallmatrix} \right)$ &  
$\left( \begin{smallmatrix} \frac{1}{9} & -\frac{\sqrt{10}}{18} \\ -\frac{\sqrt{10}}{18} & \frac{5}{18} \\ \end{smallmatrix} \right)$ &
$\left( \begin{smallmatrix} \frac{5}{18} & -\frac{\sqrt{10}}{18} \\ -\frac{\sqrt{10}}{18} & \frac{1}{9} \\ \end{smallmatrix} \right)$ &
$\left( \begin{smallmatrix} \frac{5}{9} & -\frac{\sqrt{10}}{18} \\ -\frac{\sqrt{10}}{18} & \frac{1}{18} \\ \end{smallmatrix} \right)$ &
$\frac{7}{18}$, $0$
\\
\hline
$\frac{5}{2}^+$ & $0$ & $(2,1)$ & $0$ & $0$ & $1$ & $0$ & $1$ \\
%\hline
$\frac{5}{2}^+$ & $1$ & $(2,1)$ & $0$ & $0$ & $\frac{2}{3}$ &
$\frac{1}{3}$ & $\frac{2}{3}$ \\
\hline
\end{tabular}
\caption{
  The coupling factors $\lambda^{\beta \gamma}_{\sigma \tau}$ to be used
  in the coupled-channel Faddeev equations.
  The $P_c^*$ channel corresponds to $\sigma = \frac{3}{2}$ and
  $\tau = \frac{1}{2}$, i.e. $\lambda_{\frac{3}{2} \frac{1}{2}}$.
  In the last column we write the $\lambda$ to be included
  in the single-channel Faddeev equation we use
  in this work, Eq.~(\ref{eq:PcD-trimer}).
  For uncoupled channels $\lambda = \lambda_{\frac{3}{2} \frac{1}{2}}$,
  while for coupled channels $\lambda$ is given by
  the eigenvalues of $\lambda^{\beta \gamma}_{\frac{3}{2} \frac{1}{2}}$.
}
\label{tab:lambda1-spin}
\end{table*}

\section{The $\bar{D}^* \bar{D}^* \Sigma_c$ System in the Unitary Limit}
\label{sec:Efimov}

Here we will consider the unitary limit of the previous set of
Faddeev equations. The unitary limit refers to the limit
in which a two-body system is bound at threshold, which
is interesting from a theoretical perspective because of 
its relation with the Efimov effect~\cite{Efimov:1970zz}.
Actually the $\bar{D}^* \Sigma_c$ system in the $P_c^*$ channel
is far from the unitary limit: its expected size
$1/\sqrt{2 \mu_{23} B_2} \sim 1.2\,{\rm fm}$ is
comparable with the typical hadronic size
($0.5-1.0\,{\rm fm}$).
Yet from a theoretical point of view the discussion about the unitary limit
is relevant because of the following reasons:
(i) the relation between the Efimov effect~\cite{Efimov:1970zz},
Thomas collapse~\cite{Thomas:1935zz} and
the requirement of three body forces to compensate
for the later~\cite{Bedaque:1998kg,Bedaque:1998km} and
(ii) the $\bar{D}^* \Sigma_c$ system might be closer to the unitary limit
for quantum numbers different than the ones of the $P_c^*$ channel.

As already seen, the eigenvalue equation of the $\bar{D}^* \bar{D}^* \Sigma_c$
system always reduces to 
\begin{eqnarray}
  a(p_1) = \lambda \, \tau(Z_{23}) \int \frac{d^3 p_2}{(2\pi)^3}\,
  B^0_{12}(\vec{p}_1, \vec{p}_2)\, a(p_2) \, ,
  \label{eq:3B-isospin}
\end{eqnarray}
with $\lambda$ the factor listed in Table \ref{tab:lambda1-spin},
which ranges from $\tfrac{1}{6}$ to $1$
for the isoscalar and isovector trimers.
Now we take the unitary limit, where we have
\begin{eqnarray}
  \tau(Z_{23}) &\to& - \frac{2 \pi}{\mu_{23}}\,\sqrt{\frac{\mu_{23}}{\mu_1}}
  \frac{1}{p_1} \, , \\
  \int \frac{d^2 p_2}{4 \pi}\,B^0_{12} &\to& -\frac{m_3}{2 p_1 p_2}\,
  \log{\left[ \frac{p_1^2 + p_2^2 + \frac{2 m}{m + m_3}\,p_1 p_2}
      {p_1^2 + p_2^2 - \frac{2 m}{m + m_3}\,p_1 p_2} \right]}
  \, , \nonumber \\
\end{eqnarray}
with $m = m(D^*)$, $m_3 = m(\Sigma_c)$ and where $\mu_{23}$ and $\mu_1$
are defined in Eqs.~(\ref{eq:mu_ij}) and (\ref{eq:mu_k}).
From this we arrive to
\begin{eqnarray}
  p_1\,a(p_1) &=& \frac{\lambda}{2\pi}\,
  \frac{m_3}{\mu_{23}}\sqrt{\frac{\mu_{23}}{\mu_1}}\,\frac{1}{p_1}\,
  \int_0^{\infty} dp_2\,p_2 a(p_2)\, \nonumber \\
  && \quad \times
  \log{\left[ \frac{p_1^2 + p_2^2 + \frac{2 m}{m + m_3}\, p_1 p_2}{p_1^2 + p_2^2 -\frac{2 m}{m + m_3}\, p_1 p_2} \right]}
  \, ,
\end{eqnarray}
which after the change of variable $p^2 a(p) = b(p)$ transforms into
\begin{eqnarray}
  b(p) &=& \frac{\lambda}{2\pi}\,
  \frac{m_3}{\mu_{23}}\sqrt{\frac{\mu_{23}}{\mu_1}}\,
  \int_0^{\infty} dx \,\frac{b(x p)}{x}\,
  \nonumber \\ && \quad \times
  \log{\left[ \frac{1 + x^2 + \frac{2 m}{m + m_3}\,x}{1 + x^2 - \frac{2 m}{m + m_3}\,x} \right]}
  \, .
\end{eqnarray}
This equation admits power-law solutions of the type $b(p) = p^s$,
in which case we end up with an eigenvalue equation for $s$
\begin{eqnarray}
  1 &=& 
  \frac{\lambda}{2\pi}\,
  \frac{m_3}{\mu_{23}}\sqrt{\frac{\mu_{23}}{\mu_1}}\,
  \int_0^{\infty} dx \,x^{s-1}\,
  \nonumber \\ && \quad \times
  \log{\left[ \frac{1 + x^2 + \frac{2 m}{m + m_3}\,x}
      {1 + x^2 - \frac{2 m}{m + m_3}\,x} \right]}
  \, .
\end{eqnarray}
This integral can be evaluated
analytically~\cite{Helfrich:2010yr,Helfrich:2011ut},
in which case we arrive at the eigenvalue equation 
\begin{eqnarray}
  1 = \lambda\,J_{\rm Efimov}(s, \alpha) \, , \label{eq:efimov}
\end{eqnarray}
where $J_{\rm Efimov}$ is given by
\begin{eqnarray}
  J_{\rm Efimov}(s, \alpha) = \frac{1}{\sin{2 \alpha}}\,\frac{2}{s}\,
  \frac{\sin{\alpha s}}{\cos{\frac{\pi}{2} s}} \, ,
\end{eqnarray}
and with the angle $\alpha$ determined as
\begin{eqnarray}
  \alpha = \arcsin{\left( \frac{1}{1 + \frac{m_3}{m}} \right)} \, ,
\end{eqnarray}
with $m$ and $m_3$ the masses of the charmed meson and baryon,
respectively.

The Efimov effect~\cite{Efimov:1970zz} happens when the eigenvalue equation,
i.e. Eq.~(\ref{eq:efimov}), admits complex solutions of
the type $s = \pm i s_0$,
which leads to a $b(p)$ that oscillates:
\begin{eqnarray}
  b(p) \propto
  \sin{\left[ s_0 \log{(\frac{p}{\Lambda_3})} + \varphi \right]} \, ,
\end{eqnarray}
with $\Lambda_3$ a momentum scale and $\varphi$ a phase.
From the form of $b(p)$ we deduce that the system is invariant
under the discrete scaling transformation $p \to e^{\pi/s_0} p$,
which for the case of the binding energy
reads as $B_3 \to e^{2 \pi/s_0} B_3$.
The condition for having the Efimov geometric spectrum is
\begin{eqnarray}
  \lambda \geq \lambda_c = \frac{\sin{2 \alpha}}{2 \alpha} \, ,
\end{eqnarray}
which for the $\bar{D}^* \bar{D}^* \Sigma_c$ system give us
$\lambda_c \simeq 0.861$.
From this we deduce that the $J=\frac{5}{2}$, $I=0$ trimer is potentially
Efimov-like, while all the others are not (but only for the set of
assumptions laid out at the beginning of Sect.\ref{sec:Faddeev}).
For the Efimov-like trimer we have $s_0 \simeq 0.363$, from which the
discrete scaling factor is $e^{\pi/s_0} \simeq 5711$ for momenta
and $e^{2 \pi/s_0} \simeq 3.262 \cdot 10^7$
for the trimer binding energy~\footnote{Notice that the discrete
  scaling factor for the $\bar{D}^* \bar{D}^* \Sigma_c$ system is
  different than in the three boson system, where the Efimov
  effect was originally proposed. In the three boson system
  $s_0 \simeq 1.00624$ and $e^{\pi/s_0} \simeq 22.7$,
  which can be easily reproduced from Eq.~(\ref{eq:efimov}) by simply setting
  $\lambda = 2$ and $m = m_3$. For $\lambda = 1$ and $m = m_3$ the factor
  for the standard heteronuclear Efimov effect with equal masses (i.e.
  a system of three particles with identical masses but only two resonantly
  particle pairs) $s_0 \simeq 0.4137$ and $e^{\pi/s_0} \simeq 1986.1$,
  is reproduced. Details on how the Efimov effect manifest
  in different systems can be consulted
  in Refs.~\cite{Hammer:2010kp,Naidon:2016dpf}}.
Even with a molecular $P_c^*$  close to the unitary limit,
the factors are too big to be realistically observed.
Thus the analysis presented here is mostly academical, except for one thing:
if the forces binding the trimer are short-ranged, the possibility of
the Efimov effect is related to the necessity of a three body
force for the system to be properly
renormalized~\cite{Bedaque:1998kg,Bedaque:1998km}.
The reason is the relation between the Efimov effect and
the Thomas collapse~\cite{Thomas:1935zz}:
that is, as the range of the two-body potential shrinks,
the trimer binding energy grows, eventually diverging.
However for the $\bar{D}^* \bar{D}^* \Sigma_c$ trimer, the observation
of this collapse requires interaction ranges that are orders of
magnitude smaller than the typical hadron size.

Finally, though the Efimov effect is absent
in the $\bar{D}^* \bar{D}^* \Sigma_c$ trimers under the assumptions
we are making, it might still happen in other heavy hadron systems.
For instance, the ${B}^* {B}^* \Sigma_c$ system might be a better candidate
for the Efimov effect because the associated discrete scaling factors
are larger: for $m= m(B^*)$, $m_3 = m(\Sigma_c)$ and $\lambda = 1$
we obtain $s_0 \simeq 0.652$ and $e^{\pi/s_0} \simeq 123.4$.

\section{Predictions}
\label{sec:PcD}

The basic building block of the three body calculation is
the $\bar{D}^* \Sigma_c$ interaction in the channel
with the quantum numbers of the $P_c^*$:
$J^P=\frac{3}{2}^-$ and $I=\frac{1}{2}$.
We will describe the $\bar{D}^* \Sigma_c$ system
in terms of a contact-range potential of the type
\begin{eqnarray}
  V(\bar{D}^* \Sigma_c) =
  C(\Lambda)\,f(\frac{k}{\Lambda})\,f(\frac{k'}{\Lambda}) \, ,
\end{eqnarray}
where $C(\Lambda)$ is a coupling constant and $f(\Lambda)$ a regulator function.
Here we will use a Gaussian regulator $f(x) = e^{-x^2}$ and
a cut-off window $\Lambda = 0.5-1.0\,{\rm GeV}$.
The regulator choice is arbitrary, where we have chosen a Gaussian regulator
mostly for convenience and compatibility
with previous works~\cite{Valderrama:2018sap,Liu:2018zzu}
(we will comment on the regulator dependence latter).
The cut-off choice corresponds to the expected momenta at which the description
of the $P_c^*$ as a $\bar{D}^* \Sigma_c$ bound state will break down.
In the molecular interpretation the $P_c^*$ is a $\bar{D}^* \Sigma_c$
bound state with a binding energy of $B_2 = 12 \pm 3\,{\rm MeV}$.
This binding energy is simply the difference $m(D^*) + m(\Sigma_c) - m(P_c^*)$,
where we take the isospin symmetric limit for the $D^*$ and $\Sigma_c$ masses
and with the uncertainty basically corresponding
to the experimental error in $m(P_c^*)$.
We can determine the strength of the coupling $C(\Lambda)$ 
from the condition of reproducing the binding energy $B_2$.
For this we use the two-body eigenvalue equation
\begin{eqnarray}
  1 + C(\Lambda)\,\int \frac{d^3 q}{(2 \pi)^3}\,
  \frac{f^2(\frac{q}{\Lambda})}{B_2 + \frac{q^2}{2 \mu_{23}}} = 0 \, ,
\end{eqnarray}
where $B_2$ is the two-body binding energy and $\mu_{23}$ the reduced mass
of the $\bar{D}^* \Sigma_c$ system.

\begin{figure}[ttt]
  \begin{center}
\includegraphics[width=8.5cm]{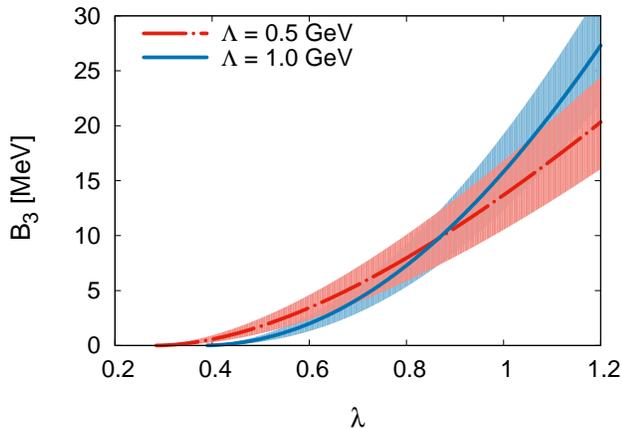}
\end{center}
\caption{
  Binding energy $B_3$ of the $\bar{D}^* \bar{D}^* \Sigma_c$ trimer
  as a function of the coefficient $\lambda$, which parametrizes
  the overlap of the trimer quantum numbers with
  the $\bar{D}^* \Sigma_c$ system
  in the $P_c^*$ channel.
  The bars represent the error coming from the uncertainty in the binding
  energy of the $P_c^*$, $B_2 = 12 \pm 3\,{\rm MeV}$.
}~\label{fig:B3-lambda}
\end{figure}

Once the contact-range coupling is obtained from the two-body eigenvalue
equation, we can solve the three-body eigenvalue equation with
$C_{\frac{3}{2} \frac{1}{2}} = C(\Lambda)$, $g(k) = f(k/\Lambda)$
and the appropriate factor $\lambda$.
Concrete calculations lead to the predictions of Table \ref{tab:trimers},
where the binding energy $B_3$ is shown for different
trimer configurations.
The binding energy $B_3$ is defined with respect to the dimer-particle
threshold, i.e. the mass of the trimers is
\begin{eqnarray}
  M = 2 m + m_3 - B_2 - B_3 \, ,
\end{eqnarray}
with $m$ and $m_3$ the mass of the $\bar{D}^*$ meson and $\Sigma_c$ baryon,
respectively.
The binding energy does not directly depend on the quantum numbers of
the trimer, but indirectly by means of the factor $\lambda$
as can be appreciated in  Table \ref{tab:trimers}.
This dependence on the coefficient $\lambda$ is shown explicitly
in Figure \ref{fig:B3-lambda}.

\begin{table}[!ttt]
\begin{tabular}{|cc|cc|c|}
\hline\hline
$J^P$ & $I$ & $B_3(\Lambda = 0.5\,{\rm GeV})$ & $B_3(\Lambda = 1.0\,{\rm GeV})$
& $\lambda$ \\ 
\hline
$\frac{1}{2}^+$ & $0$ & $4.8^{+1.5}_{-1.4}$ & $3.1^{+1.1}_{-1.0}$ & $\frac{2}{3}$ \\
$\frac{1}{2}^+$ & $1$ & $2.6^{+1.0}_{-0.9}$ & $1.3^{+0.6}_{-0.5}$ & $\frac{5}{9}$ \\
\hline
$\frac{3}{2}^+$ & $0$ & - & - & $\frac{1}{6}$ \\
$\frac{3}{2}^+$ & $1$ & $0.5^{+0.3}_{-0.2}$ & - & $\frac{7}{18}$ \\
\hline
$\frac{5}{2}^+$ & $0$ & $14 \pm 3$ & $16^{+3}_{-4}$ & $1$ \\
$\frac{5}{2}^+$ & $1$ & $4.8^{+1.5}_{-1.4}$ & $3.1^{+1.1}_{-1.0}$ & $\frac{2}{3}$ \\
  \hline\hline 
\end{tabular}
\caption{
  Predictions for the binding energy $B_3$ (in units of MeV)
  of the $\bar{D}^* \bar{D}^* \Sigma_c$
  trimers for different quantum numbers and for a cut-off
  $\Lambda = 0.5-1.0\,{\rm GeV}$, where $B_3$ is relative
  to the $\bar{D}^* P_c^*$ threshold (i.e. $B_3 > 0$ indicates
  that the trimer binds).
  The errors in $B_3$ are a consequence of
  the uncertainty in the $\bar{D}^* \Sigma_c$ binding energy,
  $B_2 = 12 \pm 3\,{\rm MeV}$.
}
\label{tab:trimers}
\end{table}

The trimer binding energies are affected by a series of uncertainties,
which we will discuss below.
The results of Table \ref{tab:trimers} already contain two error sources,
the binding energy of the $P_c^*$ ($B_2 = 12 \pm 3\,{\rm MeV}$)
and the cut-off window ($\Lambda = 0.5-1.0\,{\rm GeV}$).
Assuming that the $P_c^*$ is indeed molecular, the next most important
source of uncertainty is the the choice of which interactions
are leading and subleading.
Previously we have assumed that the only leading order interaction
is the $\bar{D}^* \Sigma_c$ short-range potential in the $P_c^*$ channel.
We will review this assumption in detail in the following lines,
where the different possibilities will be named scenarios A, B and C.

We begin with the $\bar{D}^* \bar{D}^*$ interaction.
T\"ornqvist pointed out~\cite{Tornqvist:1993ng} that the flavour exotic
configurations of this system are less likely to bind in general.
But this conclusion is probably incomplete because it relies on one-pion
exchange while ignoring the short-range contributions to
the $\bar{D}^* \bar{D}^*$ interaction.
In this regard several works~\cite{Carlson:1987hh,Gelman:2002wf,Vijande:2009kj,Junnarkar:2018twb} (\cite{Karliner:2017qjm,Eichten:2017ffp,Mehen:2017nrh})
have indicated the possibility of a isoscalar flavour exotic $1^+$ tetraquark
below (above) the $\bar{D} \bar{D}^*$ threshold.
If close enough to the $\bar{D}^* \bar{D}^*$ threshold, it might contribute
to the dynamics of this two-body system, suggesting a strong attraction
in the $J=1$, $I=0$ channel.
In turns out that this contribution will reduce/increase the binding of
the isovector $J^P = \tfrac{1}{2}^+$/$\tfrac{3}{2}^+$ trimer.
The reason for this reduction/increase of the binding energy are
the $\lambda^{\beta \gamma}_{\sigma \tau}$ coefficients
in Table \ref{tab:lambda1-spin}:
for the isovector $J^P = \tfrac{1}{2}^+$/$\tfrac{3}{2}^+$ three-body state,
attraction in the isoscalar $\bar{D}^* \bar{D}^*$ subsystem
forces the trimer into a configuration that has less/more
overlap with the $P_c^*$ channel of the $\bar{D}^* \Sigma_c^*$ subsystem.
Scenario A will represent the possibility of a shallow isoscalar
$J=1$ $\bar{D}^* \bar{D}^*$ bound state located at threshold,
where predictions for the binding energy of the two affected
trimers can be found in Table \ref{tab:trimer-landscape}.

Next we consider the $\bar{D}^* \Sigma_c$ interaction in channels
different than the $P_c^*$ ($I=\tfrac{1}{2}$, $J^P = \frac{3}{2}^-$).
Phenomenology, in particular the hidden-gauge model~\cite{Wu:2010jy},
indicates that the interaction in the $I=\tfrac{1}{2}$,
$J^P = \tfrac{1}{2}^-$ channel could very well be
as attractive as in the $P_c^*$ channel.
Scenario B will refer to this possibility.
If this is the case the three isoscalar trimers will be degenerate,
having the same binding energy as the isoscalar
$J^P = \tfrac{5}{2}^+$ trimer, see Table \ref{tab:trimer-landscape}.
The $\bar{D}^* \Sigma_c^*$ interaction in the $I=\tfrac{3}{2}$ channel
has received less attention, though it could also be
strong and attractive~\cite{Chen:2015loa}.
Scenario C will consider that there is a $I=\tfrac{3}{2}$
$\bar{D}^* \Sigma_c^*$ bound state at threshold for both
$J^P = \tfrac{1}{2}^-$ and $\tfrac{3}{2}^-$ (while also
assuming scenario B).
In this scenario the isovector trimers will be degenerate and
more bound than expected, see Table \ref{tab:trimer-landscape}.
The degeneration is again a consequence of
the $\lambda^{\beta \gamma}_{\sigma \tau}$ coefficients of
Table \ref{tab:lambda1-spin}: the isovector trimers
can always be in a configuration with the same relative contribution
from the $I=\tfrac{1}{2}$ and $\tfrac{3}{2}$ $\bar{D}^* \Sigma_c^*$ channels.
In addition the hypothesis of $\bar{D}^* \Sigma_c^*$ spin degeneracy
in scenario B means that the trimers are effectively decoupled of
the isoscalar $J=1$ $\bar{D}^* \bar{D}^*$ interaction.
That is, the combination of scenarios A and C leads to
the same predictions as scenario C alone, see
again Table \ref{tab:trimer-landscape}.

Another aspect of the $\bar{D}^* \Sigma_c$ interaction is
its long-range piece, which is given by one-pion exchange (OPE).
Here we have considered OPE to be subleading.
This assumption can be analyzed with the formalism of Refs.~\cite{Valderrama:2012jv,Lu:2017dvm},
which investigated the range of momenta for which OPE is
perturbative in the heavy meson-meson and heavy baryon-baryon systems.
By adapting the methods of Refs.~\cite{Valderrama:2012jv,Lu:2017dvm} to
the heavy meson-baryon case, i.e. to $\bar{D}^* \Sigma_c$,
we arrive at the conclusion that in the $P_c^*$ channel
OPE is perturbative for $p < \Lambda_{\rm OPE} \simeq 270-450\,{\rm MeV}$
in the chiral limit ($m_{\pi} = 0$, with $m_{\pi}$ the pion mass).
The uncertainty is a consequence of the axial coupling constant of the pion
with the charmed meson $\Sigma_c$, which is not known experimentally,
see Ref.~\cite{Cheng:2015naa} for a recent summary.
For the physical pion mass, $m_{\pi} \simeq 140\,{\rm MeV}$,
we expect $\Lambda_{\rm OPE} \simeq 410-710\,{\rm MeV}$, see 
Refs.~\cite{Valderrama:2012jv,Lu:2017dvm} for a detailed explanation.
If we take into account that the binding momentum of a molecular $P_c^*$
is about $160\,{\rm MeV}$, it is natural to expect OPE to be subleading,
although it will provide important corrections to
the leading order description.
The binding momentum of the $\tfrac{5}{2}^+$ trimer, the most
bound three-body state we are predicting, is of the same order:
$\sqrt{2 \mu_1 B_3} \sim 200-210\,{\rm MeV}$
(or, in terms of size $1 / \sqrt{2 \mu_1 B_3} \sim 0.9-1.0\,{\rm fm}$).
Owing to the similarity in the scales involved, we expect the conclusions
derived in the two-body sector to apply in the three-body sector,
in agreement with our original assumption.

\begin{table}[!ttt]
\begin{tabular}{|c|cc|c|}
\hline\hline
Scenario & $J^P$ & $I$ & $B_3$ \\ 
\hline
A & $\frac{1}{2}^+$ & $1$ & $0.5-1.9$ \\
A & $\frac{3}{2}^+$ & $1$ & $0.1-0.9$ \\
\hline
B & $\frac{1}{2}^+$ & $0$ & $14-16$ \\
B & $\frac{3}{2}^+$ & $0$ & $14-16$ \\
\hline
C/A+C & $\frac{1}{2}^+$ & $1$ & $6.7-7.3$ \\
C/A+C & $\frac{3}{2}^+$ & $1$ & $6.7-7.3$ \\
C/A+C & $\frac{5}{2}^+$ & $1$ & $6.7-7.3$ \\
\hline\hline
\end{tabular}
\caption{
  Different scenarios for the predictions for the binding energy $B_3$ of
  the $\bar{D}^* \bar{D}^* \Sigma_c$ trimers.
  Scenario A refers to the existence of a $\bar{D}^* \bar{D}^*$ bound state
  at threshold. Scenario B is when the $I=\frac{1}{2}$ and
  $J^P = \frac{1}{2}^-$ and $\frac{3}{2}^-$ $\bar{D}^* {\Sigma}_c^*$
  are degenerate.
  Scenario C assumes that the $I=\frac{3}{2}$ $\bar{D}^* {\Sigma}_c^*$
  interaction is strong, generating a bound state at threshold for
  both $J^P = \frac{1}{2}^-$ and $\frac{3}{2}^-$.
  For each scenario only the quantum numbers affected are shown.
}
\label{tab:trimer-landscape}
\end{table}

Finally, though the isoscalar $\tfrac{5}{2}^+$ trimer depends only
on the interaction in the $P_c^*$ channel, it actually contains
an additional source of uncertainty.
The Efimov effect~\cite{Efimov:1970zz} can happen in this trimer, as shown by
the analysis of the $\bar{D}^* \bar{D}^* \Sigma_c$ system
in the unitary limit, see Sect.~\ref{sec:Efimov}.
The $\bar{D}^* \Sigma_c$ two-body system in the $P_c^*$ channel
is actually far from the unitary limit, but the analysis is still
relevant because of the relation between the Efimov effect
and Thomas collapse~\cite{Thomas:1935zz}.
The idea is that, even though a molecular $P_c^*$ is not in the unitary limit,
for $\Lambda \to \infty$ the isoscalar $\tfrac{5}{2}^+$ trimer will
collapse, i.e. its binding energy will diverge ($B_3 \to \infty$).
This collapse can be prevented with the inclusion of a three-body force,
without which the trimer binding energy predictions
will not be formally cut-off independent~\cite{Bedaque:1998kg,Bedaque:1998km}.
In practice, owing the large discrete scaling factor of $e^{\pi/s_0} \simeq 5711$,
the divergence of the trimer binding requires fantastically
large cut-offs to be noticed.
For instance, the cut-off required for the first nonphysical
isovector $J^P = \tfrac{5}{2}^+$ trimer to appear
is $\Lambda \sim 62\,{\rm GeV}$, which is pretty large.
Thus it is not surprising that the cut-off uncertainty for this trimer
is not particularly big, about $2\,{\rm MeV}$ in the
$\Lambda = 0.5-1.0\,{\rm GeV}$ cut-off window.
Alternatively we can check the model dependence of
the $J^P = \tfrac{5}{2}^+$ trimer prediction by using a different regulator,
for instance a delta-shell in coordinate space
\begin{eqnarray}
  V(r; \bar{D}^* \Sigma_c) =
  C(R_C)\,\frac{\delta(r - R_c)}{4 \pi R_c^2} \, ,
\end{eqnarray}
where $R_c$ is a cut-off radius which we take of the order of the typical
hadronic size: $R_c = 0.5-1.0\,{\rm fm}$.
The delta-shell potential is actually a separable interaction,
where its momentum space form is
\begin{eqnarray}
  V(\bar{D}^* \Sigma_c) =
  C(R_C)\,\frac{\sin(k R_c)}{k R_c}\,\frac{\sin(k' R_c)}{k' R_c} \, .
\end{eqnarray}
With this regulator the prediction for the location of the $\tfrac{5}{2}^+$
trimer binding is $B_3 = 14 \pm 3\,{\rm MeV}$ ($17^{+3}_{-4}\,{\rm MeV}$)
for $R_c = 1.0\,{\rm fm}$ ($0.5\,{\rm fm}$), i.e. almost identical
to the Gaussian regulator predictions.
Despite the previous checks caution is advised
because the formal requirement of a three-body force,
even if the related divergence happens at really large cut-offs,
indicates the existence of systematic uncertainties
that are not being taken into account.

\section{Conclusions}
\label{sec:Conclusions}

The hypothesis that the $P_c^*$ is a $I=\tfrac{1}{2}$,
$J^P = \tfrac{3}{2}^{-}$ $\bar{D}^* \Sigma_c$ molecule
implies the existence of a few $\bar{D}^* \bar{D}^* \Sigma_c$ trimers.
Calculations in a contact-range theory indicate that the most bound
of these trimers has the quantum numbers $I=0$, $J^P = \tfrac{5}{2}^+$
and a binding energy $B_3 \sim 14-16\,{\rm MeV}$.
There are other three or four trimer configurations that are likely to bind,
but they are expected to be considerably less bound: two states
with $B_3 \sim 3-5\,{\rm MeV}$ with quantum numbers
$I=0$, $J^P = \frac{1}{2}^+$ and $I=1$, $J^P = \frac{5}{2}^+$,
a state with $B_3 \sim 1-3\,{\rm MeV}$ and quantum numbers
$I=1$, $J^P = \frac{1}{2}^+$ and maybe a state on the verge
of binding with $I=1$, $J^P = \frac{3}{2}^+$,
see Table \ref{tab:trimers} for details.
These predictions are affected by a series of uncertainties,
which mostly stem from the fact that we do not know too much about
the $\bar{D}^* \Sigma_c$ interaction except in the $P_c^*$ channel
(and even this is dependent on the nature of the $P_c^*$).
These uncertainties are taken into account on the basis of considering
different hypothesis about the $\bar{D}^* \bar{D}^*$ and
$\bar{D}^* \Sigma_c$ interactions, which are summarized
in Table \ref{tab:trimer-landscape}.
From the previous considerations we arrive to the conclusion
that  the most solid prediction is that of the isoscalar
$J^P = \frac{5}{2}^+$ trimer.

We have also considered the $\bar{D}^* \bar{D}^* \Sigma_c$ system in the unitary
limit, i.e. when the $P_c^*$ is located at the $\bar{D}^* \Sigma_c$ threshold.
In this situation the $J^P = \frac{5}{2}^+$ trimer will be Efimov-like and
will have a geometric spectrum with a scaling factor of $3.3 \cdot 10^7$
for the binding energies.
This scaling factor is fantastically large, which implies
that there would be no practical way to observe it even if
we could tune the $\bar{D}^* \Sigma_c$ scattering length.
Yet the possibility of the Efimov effect in the isoscalar
$J^P = \frac{5}{2}^+$ trimer implies that a three-body force
should be included in the calculations at leading order.
The practical importance of this three-body force might be tangential,
as reflected by the mild cut-off dependence of the isoscalar
$J^P = \frac{5}{2}^+$ trimer binding in the cut-off window
chosen in this work, $\Lambda = 0.5-1.0\,{\rm GeV}$.

Without knowing the nature of the $P_c^*$,
the predictions of this work will remain theoretical:
if the $P_c^*$ is not molecular but a compact pentaquark instead
the $\bar{D}^* \bar{D}^* \Sigma_c$ trimers are not expected to bind.
The experimental production of these trimers is expected to be difficult,
with the lattice probably providing a more convenient way to investigate them.
Finally we notice that the existence of $\Sigma_c \Sigma_c \bar{D}^*$ bound
trimers is also likely, but their location will be subjected to even larger
uncertainties as a consequence of the $\Sigma_c \Sigma_c$ interaction,
which is not particularly well-known but probably strong.

\section*{Acknowledgments}
	
I would like to thank Li-Sheng Geng, Kanchan P. Khemchandani and
Alberto Martinez Torres for discussions and a critical and
careful reading of this manuscript.
This work is partly supported by the National Natural Science Foundation
of China under Grants No.11522539, No.11735003, the Fundamental
Research Funds for the Central Universities and the Thousand
Talents Plan for Young Professionals.

%\bibliography{PcD.bib}

\end{document}